\documentclass{ws-ijmpe}

\usepackage{graphicx}
\begin{document}

\markboth{Authors' Names}{Constraining the Symmetry Energy}
\catchline{}{}{}{}{}

\title{CONSTRAINING THE SYMMETRY ENERGY: A JOURNEY IN THE ISOSPIN PHYSICS 
FROM COULOMB BARRIER TO 
DECONFINEMENT}
\author{M. DI TORO $^*$, M. COLONNA, V. GRECO, G. FERINI, C. RIZZO, J. RIZZO}%
\address{Laboratori Nazionali del Sud INFN, I-95123 
Catania, Italy,\\
and Physics-Astronomy Dept., University of Catania\\%
$^*$ E-mail:\_ditoro@lns.infn.it}

\author{V. BARAN}%
\address{Dept.of Theoretical Physics, Bucharest Univ., and NIPNE-HH,
Magurele, Bucharest, Romania}

\author{T. GAITANOS}%
\address{Inst. f\"ur Theoretische Physik, Universit\"at Giessen, 
D-35312 Giessen, Germany}

\author{V. PRASSA}%
\address{Dept.of Theoretical Physics, Aristotle Univ. of Thessaloniki,
Gr-54124, Greece}

\author{H. H. WOLTER}%
\address{Dept. f\"ur Physik, Universit\"at M\"unchen, 
D-85748 Garching, Germany}

\author{M. ZIELINSKA-PFABE}%
\address{Smith College, Northampton, Mass. USA}

\maketitle
\begin{history}%
\received{October 07}%
\revised{October 07}%
\end{history}

\begin{abstract}
Heavy Ion Collisions ($HIC$) represent a unique tool to probe the in-medium
nuclear interaction in regions away from saturation. In this work we present a 
selection of reaction observables in dissipative collisions 
particularly sensitive to the isovector part of the interaction, i.e. to the
symmetry term of the nuclear Equation of State ($EoS$).
At low energies the behavior of the symmetry energy around saturation 
influences dissipation and fragment production mechanisms.
We will first discuss the recently observed Dynamical Dipole Radiation, due
to a collective neutron-proton oscillation during the charge equilibration
in fusion and deep-inelastic collisions. 
Important $Iso-EOS$ are stressed. Reactions induced by unstable $^{132}Sn$ 
beams appear
to be very promising tools to test the sub-saturation Isovector $EoS$.
New Isospin sensitive observables 
are also presented for deep-inelastic, fragmentation collisions and Isospin
equilibration measurements (Imbalance Ratios).  

The high density symmetry term can be derived from
 isospin effects on heavy ion reactions
at relativistic energies (few $AGeV$ range), that can even allow
a ``direct'' study of the covariant structure of the isovector interaction
in the hadron medium. 
Rather sensitive observables are proposed from collective flows
and from pion/kaon production. 
The possibility of the transition to a mixed hadron-quark phase, 
at high baryon and isospin density, is finally suggested. Some signatures
could come from an expected ``neutron trapping'' effect.
The importance of studying violent collisions with radioactive beams
from low to relativistic energies is finally stressed.

\end{abstract}


\vskip -1.0cm
\section{Introduction}

The symmetry energy $E_{sym}$ appears in the energy density
$\epsilon(\rho,\rho_3) \equiv \epsilon(\rho)+\rho E_{sym} (\rho_3/\rho)^2
 + O(\rho_3/\rho)^4 +..$, expressed in terms of total ($\rho=\rho_p+\rho_n$)
 and isospin ($\rho_3=\rho_p-\rho_n$) densities. The symmetry term gets a
kinetic contribution directly from basic Pauli correlations and a potential
part from the highly controversial isospin dependence of the effective 
interactions \cite{baranPR}. Both at sub-saturation and supra-saturation
densities, predictions based of the existing many-body techniques diverge 
rather widely, see \cite{fuchswci}. 
We remind that the knowledge of the 
$EoS$ of asymmetric matter is very important at low densities (neutron skins,
nuclear structure at the drip lines, neutron distillation in fragmentation,
 neutron star formation and crust..) as well as at high densities (transition 
to a deconfined phase, neutron star mass/radius, cooling, hybrid structure,
 formation of black holes...). 
We take advantage of new opportunities in 
theory (development of rather reliable microscopic transport codes for $HIC$)
 and in experiments (availability of very asymmetric radioactive beams, 
improved possibility of measuring event-by-event correlations) to present
results that are severely constraining the existing effective interaction 
models. We will discuss dissipative collisions in a wide range of energies, 
 from just above the Coulomb barrier up to a few $AGeV$. 
Low to Fermi energies
 will bring information on the symmetry term around (below) normal density, 
relativistic energies will probe high density regions, even testing the
covariant structure of the isovector terms. 
The transport codes are based on 
mean field theories, with correlations included via hard nucleon-nucleon
elastic and inelastic collisions and via stochastic forces, selfconsistently
evaluated from the mean phase-space trajectory, see 
\cite{baranPR,guarneraPLB373,colonnaNPA642,chomazPR}. Stochasticity is 
essential in 
order to get distributions as well as to allow the growth of dynamical 
instabilities.  
 The isovector part of the $EoS$ has been tested systematically by using two 
different behaviors of the symmetry energy below saturation: 
one ($Asysoft$) where it is a smooth decreasing function towards low densities,
 and another one ($Asystiff$) where we have a rapid decrease, 
\cite{baranPR,colonnaPRC57}.

\section{The Prompt Dipole $\gamma$-Ray Emission}

The possibility of an entrance channel bremsstrahlung dipole radiation
due to an initial different N/Z distribution was suggested at the beginning
of the nineties \cite{ChomazNPA563,BortignonNPA583}. 
After several experimental evidences, in fusion as well as in deep-inelastic
reactions \cite{FlibPRL77,CinNC111,PierrouEPJA16,AmoPRC29,PierrouPRC71,medea} 
we have
now a good understanding of the process and stimulating new perspectives
from the use of radioactive beams.

During the charge equilibration process taking place
 in the first stages of dissipative reactions between colliding ions with
 different N/Z
ratios, a large amplitude dipole collective motion develops in the composite
dinuclear system, the so-called dynamical dipole mode. This collective dipole
gives rise to a prompt $\gamma $-ray emission which depends:
 i) on the absolute
value of the intial dipole moment
\begin{eqnarray}
&&D(t= 0)= \frac{NZ}{A} \left|{R_{Z}}(t=0)- {R_{N}}(t=0)\right| =  \nonumber \\
&&\frac{R_{P}+R_{T}}{A}Z_{P}Z_{T}\left| (\frac{N}{Z})_{T}-(\frac{N}{Z})_{P}
\right|,
\label{indip}
\end{eqnarray}
being ${R_{Z}}= \frac {\Sigma_i x_i(p)}{Z}$ and
${R_{N}}=\frac {\Sigma_i x_i(n)}{N} $ the
center of mass of protons and of neutrons respectively, while R$_{P}$ and
R$_{T}$ are the
projectile and target radii; ii) on the fusion/deep-inelastic dynamics;
 iii) on the symmetry term, below saturation, that is acting as a restoring
force.

A detailed description is obtained in a
microscopic approach based on semiclassical transport equations,
of Landau-Vlasov type, \cite{BrinkNPA372},
where mean field and two-body collisions are treated in a
self-consistent way, see details in \cite{BaranNPA600,BaranNPA679}. Realistic
effective interactions of Skyrme type are used. 
The
resulting physical picture is in good agreement with quantum
Time-Dependent-Hartree-Fock calculation \cite{SimenPRL86}. In
particular we can study in detail how a collective dipole
oscillation develops in the entrance channel \cite{BaranNPA679}.

We can follow the time evolution
of the dipole moment
in the $r$-space,
 $D(t)= \frac{NZ}{A} ({R_{Z}}- {R_{N}})$ and in
$p-$space, $DK(t)=(\frac{P_{p}}{Z}-\frac{P_{n}}{N})$, 
with $P_{p}$
($P_{n}$) center of mass in momentum space for protons (neutrons),
is just the canonically conjugate momentum of the $D(t)$ coordinate,
i.e. as operators $[D(t),DK(t)]=i\hbar$
see \cite{BaranNPA679,SimenPRL86,BaranPRL87}. 
A nice "spiral-correlation"
clearly denotes the collective nature
 of the mode.

From the dipole evolution given from the Landau-Vlasov transport
we can directly
apply a bremsstrahlung {\it ("bremss")}  approach
\cite{BaranPRL87} to estimate the ``direct'' photon emission probability
($E_{\gamma}= \hbar \omega$):
\begin{equation}
\frac{dP}{dE_{\gamma}}= \frac{2 e^2}{3\pi \hbar c^3 E_{\gamma}}
 |D''(\omega)|^{2}  \label{brems},
\end{equation}
where $D''(\omega)$ is the Fourier transform of the dipole acceleration
$D''(t)$. We remark that in this way it is possible
to evaluate, in {\it absolute} values, the corresponding pre-equilibrium
photon emission.

We must add a
couple of comments of interest for the experimental selection of the Dynamical
Dipole: i) The centroid is always shifted to lower energies (large
deformation of the dinucleus); ii) A clear angular anisotropy should be present
since the prompt mode has a definite axis of oscillation
(on the reaction plane) at variance with the statistical $GDR$.
In a very recent experiment the prompt dipole radiation has been 
investigated with
a $4 \pi$ gamma detector. A strong dipole-like photon angular distribution
$(\theta_\gamma)=W_0[1+a_2P_2(cos \theta_\gamma)]$, $\theta_\gamma$ being the 
angle between the emitted photon and the beam axis, has been observed,
 with the 
$a_2$ parameter close to $-1$, see \cite{medea}. 
\begin{figure}[htb]
\begin{center}
\centering
%\begin{picture}(0,0)
%\put(120.0,4.8){\mbox{\includegraphics[width=4.0cm]{emiss.ps}}}
%\end{picture}
\includegraphics[scale=0.30]{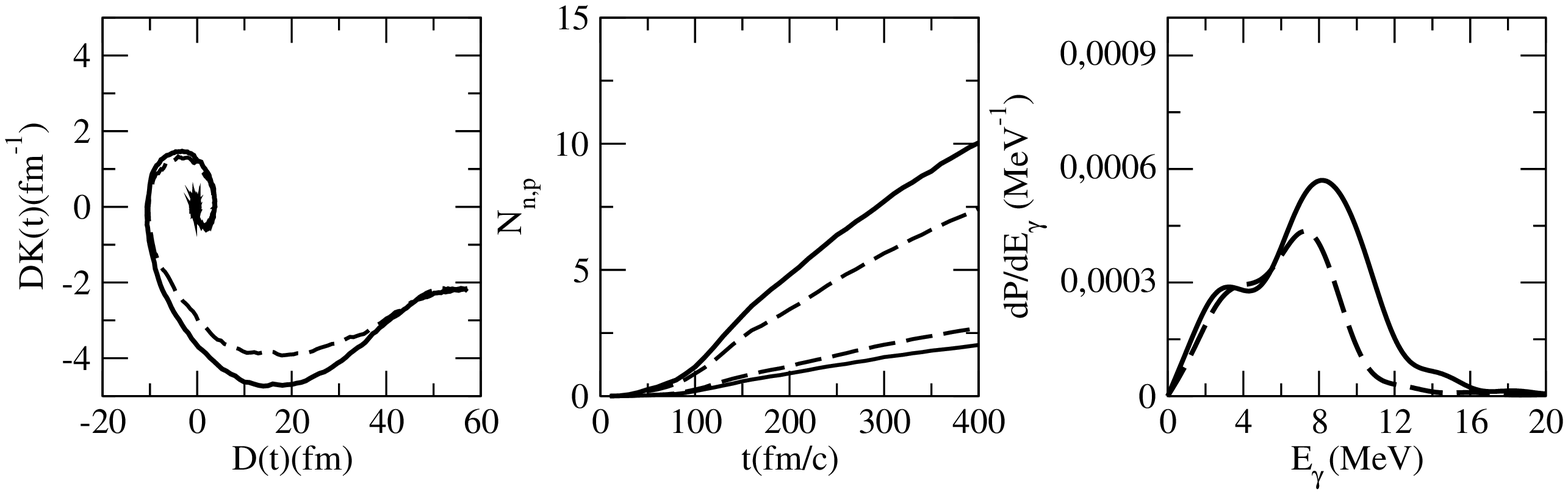}
\vskip 1.5cm
\includegraphics[scale=0.30]{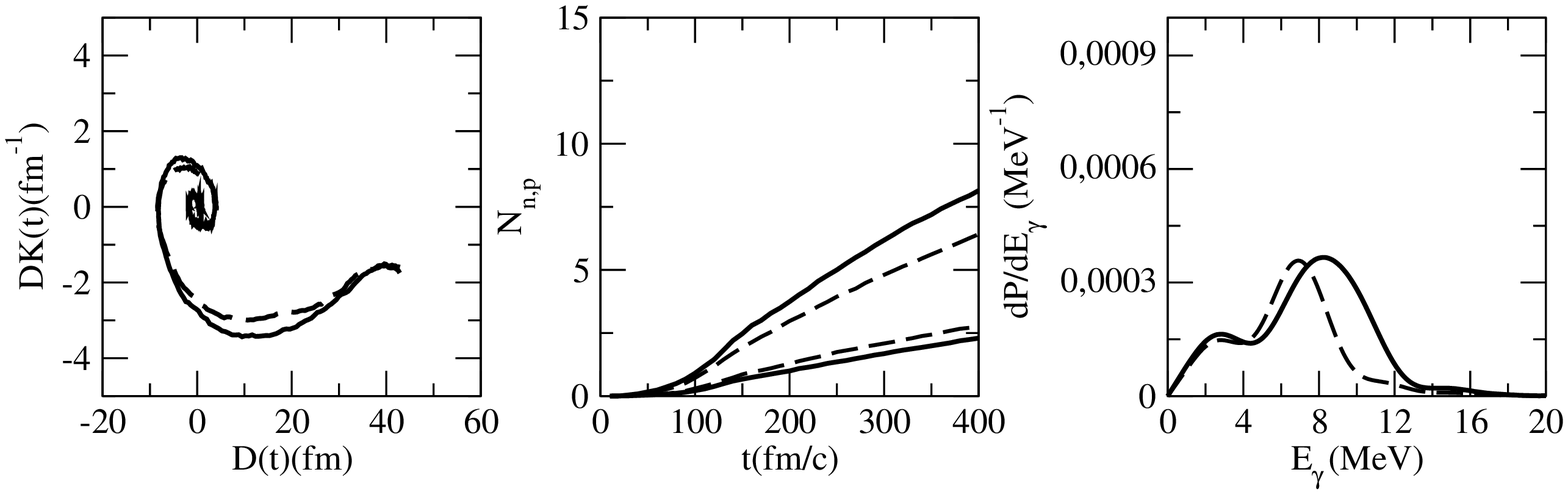}
%\vskip -1.0cm
\caption{Upper Curves: $^{132}$Sn +$^{58}$Ni system ($E=10 AMeV$, $b=4fm$).
 Lower Curves: same reactions but induced by $^{124}$Sn.
Left Panel: DK-D Spirals. Central panel: neutron (upper) and proton (lower) 
emissions. Right panel: $\gamma$ spectrum. Asy-soft (solid lines) and asy-stiff
(dashed lines) symmetry energies.}
\end{center}
\label{sn132}
\end{figure}

The deviation from a $pure$
dipole form can be interpreted as due to the rotation of the dinucleus symmetry
axis vs. the beam axis during the Prompt Dipole Emission. From accurate 
angular 
distribution mesurements we can then expect to get a direct information on the
Dynamical Dipole Life Time.

At higher beam energies we expect a decrease of the direct dipole
radiation for two main reasons both due to the increasing importance of hard
NN collisions: i) a larger fast nucleon emission that will equilibrate the
isospin before the collective dipole starts up; ii) a larger damping of the
collective mode due to $np$ collisions.

The prompt dipole radiation also represents a nice cooling mechanism on the
fusion path. It could be a way to pass from a {\it warm} to a {\it cold}
fusion in the synthesis of heavy elements with a noticeable increase of the
{\it survival} probability, \cite{luca04},

\begin{equation}
\frac{P_{surv,dipole}}{P_{surv}}=\frac{P_\gamma P_{surv}(E^*-E_\gamma)}
{P_{surv}(E^*)} + (1-P_\gamma) > 1, \frac{P_{surv}(E^*-E_\gamma)}
{P_{surv}(E^*)} > 1
\label{surv}
\end{equation}

\subsubsection*{Symmetry Energy Effects}

The use of unstable neutron rich projectiles would largely increase the
effect, due to the possibility of larger entrance channel asymmetries.
In order to suggest proposals for the new $RIB$ facility $Spiral~2$, 
\cite{lewrio} we have studied fusion events in the reaction $^{132}Sn+^{58}Ni$ 
at $10AMeV$, \cite{spiral2}. We espect a $Monster$ Dynamical Dipole, 
the initial 
dipole moment $D(t=0)$ being of the order of 50fm, about two times the largest 
values probed so far, allowing a detailed study of the symmetry potential, 
below 
saturation,
responsible of the restoring force of the dipole oscillation and even 
of the damping,
 via the fast neutron emission. 
In the Fig.1 (upper) we present some preliminary very 
promising results.
The larger value of the symmetry energy for the $Asysoft$ choice at 
low densities,
where the prompt dipole oscillation takes place, leads to some clear 
observable 
effects:
i) Larger Yields, as we see from the larger amplitude of the "Spiral" 
(left panel) 
and
finally in the spectra (right panel); ii) Larger mean gamma energies, 
shift of the 
centroid
to higher values in the spectral distribution (right panel); iii) Larger 
width of the
"resonance" (right panel) due to the larger fast neutron emission (central 
panel).
We note the opposite effect of the Asy-stiffness on neutron vs proton 
emissions.
The latter point is important even for the possibility of an independent 
test just
measuring the $N/Z$ of the pre-equilibrium nucleon emission. 

The symmetry energy influence can be of the order of $20\%$, and so well 
detected.
In the lower part of the same figure we present the same results for 
reactions induced by
the stable $^{124}Sn$ beam. We still se the $Iso-EoS$ effects, but 
largely reduced.

\vskip -1.0cm
\section{Isospin Dynamics in Neck Fragmentation at Fermi Energies}

It is now quite well established that the largest part of the reaction
cross section for dissipative collisions at Fermi energies goes
through the {\it Neck Fragmentation} channel, with $IMF$s directly
produced in the interacting zone in semiperipheral collisions on very short
time scales \cite{colonnaNPA589,wcineck}. We can predict interesting 
isospin transport 
effects for this new
fragmentation mechanism since clusters are formed still in a dilute
asymmetric matter but always in contact with the regions of the
projectile-like and target-like remnants almost at normal densities.
Since the difference between local neutron-proton chemical potentials is given 
by $\mu_n-\mu_p=4E_{sym}(\rho_3/\rho)$, we expect a larger neutron flow to
 the neck clusters for a stiffer symmetry energy around saturation, 
\cite{baranPR,baranPRC72}. The isospin dynamics can be directly extracted 
from correlations between $N/Z$, $alignement$ and emission times of the $IMF$s.
The alignment between $PLF-IMF$ and $PLF-TLF$ directions
represents a very convincing evidence of the dynamical origin of the 
mid-rapidity fragments produced on short time scales \cite{baranNPA730}. 
The form of the
$\Phi_{plane}$ distributions (centroid and width) can give a direct
information on the fragmentation mechanism \cite{dynfiss05}. Recent 
calculations confirm that the light fragments are emitted first, a general 
feature expected for that rupture mechanism \cite{liontiPLB625}. 
The same conclusion can be derived from {\it direct} emission time 
measurements based on deviations from Viola systematics  observed
in event-by-event velocity correlations between $IMF$s and the $PLF/TLF$ 
residues
 \cite{baranNPA730,dynfiss05,velcorr04}. 
 We can figure out
   a continuous transition from fast produced fragments via neck instabilities
   to clusters formed in a dynamical fission of the projectile(target) 
   residues up to the evaporated ones (statistical fission). Along this 
   line it would be even possible to disentangle the effects of volume
   and shape instabilities. 
A neutron enrichment of the overlap ("neck") region is
   expected, due to the neutron migration from higher (spectator) to 
   lower (neck) density regions, directly related to
%connected to 
   the slope of the symmetry energy \cite{liontiPLB625}. 
%Neutron and/or 
%light isobar measurements
%   in different rapidity regions appear important
%\cite{milazzo,colinPRC67}.
%%%%%%%%%%%%%%%%%%%%%%%%
%\begin{figure}
%%\begin{center}
% \includegraphics[scale=0.30]{nzphiplane.eps}
%% \includegraphics[scale=0.3,angle=-90]{nzphicor.eps}
%\begin{picture}(0,0)
%\put(3.,170.){\mbox{\includegraphics[scale=0.32,angle=-90]{nzphicor.eps}}}
%\end{picture}
%\vskip -1.0cm
%\caption{Correlation between $N/Z$ of $IMF$ and $alignement$ in ternary
%events of  the $^{132}Sn+^{64}Ni$
%reaction at $35~AMeV$. $Left~panel$. Exp. results: points correspond to fast 
%formed $IMF$s (Viola-violation selection); histogram for all $IMF$s at 
%mid-rapidity (including statistical emissions). $Right~Panel$. Simulation 
%results: squares, asysoft; circles, asystiff}  
%\vskip -1.0cm
%\label{nzphi}
%\end{center}
%\end{figure}
%%%%%%%%%%%%%%%%%%%%%%%%
%%%%%%%%%%%%%%%%%%%%%%%%
%\begin{figure}
%\begin{center}
% \includegraphics[scale=0.25]{nzphiplane.eps}
%\includegraphics[angle=-90,scale=0.30]{nzphicor.eps}
%\vskip -1.0cm
%\caption{Correlation between $N/Z$ of $IMF$ and $alignement$ in ternary
%events of  the $^{132}Sn+^{64}Ni$
%reaction at $35~AMeV$. $Left~panel$. Exp. results: points correspond to fast 
%formed $IMF$s (Viola-violation selection); histogram for all $IMF$s at 
%mid-rapidity (including statistical emissions). $Right~Panel$. Simulation 
%results: squares, asysoft; circles, asystiff}  
%\vskip -1.3cm
%\label{nzphi}
%\end{center}
%\end{figure}
%%%%%%%%%%%%%%%%%%%%%%%%
A very nice new analysis has been presented on the $Sn+Ni$ data at $35~AMeV$
by the Chimera Collab., Fig.2 of ref.\cite{defilposter}.
%, see Fig.\ref{nzphi} left panel.
A strong correlation between neutron enrichemnt and alignement (when the 
short emission time selection is enforced) is seen, that can be reproduced 
only with 
a stiff behavior of the symmetry energy. {\it This is the 
first clear evidence in favor of a relatively large slope (symmetry pressure) 
around saturation}.

\vskip -1.0cm
\section{Isospin Distillation with Radial Flow}

In central collisions at 30-50 MeV/A, where the full disassembly of the system
into many fragments is observed, one can study specifically properties of
liquid-gas phase transitions occurring in asymmetric matter 
\cite{mue95,bao197,chomazPR,baranPR}. 
For instance,
in neutron-rich matter, phase co-existence leads to
a different asymmetry
in the liquid and gaseous phase:  fragments (liquid) appear more symmetric
with respect to the initial matter, while light particles (gas) are
more neutron-rich.
The amplitude of this effect
%, qualitatively related to the decrease of
%%fact that
%the symmetry energy when the density gets lower,
depends on
 specific properties of the isovector part of the nuclear interaction,
namely on the value and the derivative of the symmetry
energy at low density.  
%, that is
%complementary to the information obtained from the investigation of the
%properties of promptly emitted particles, the so-called pre-equilibrium
%emission.

This investigation is interesting in a more general context:
In heavy ion collisions the dilute phase appears during the expansion
of the interacting matter.
%and so we expect interesting observables related
%to the coupling of distillation and expansion dynamics.
Thus we study effects of the coupling of expansion, fragmentation and 
distillation in a 
two-component (neutron-proton) system.
% in a dynamical model.
We present fully consistent dynamical results, based on stochastic
microscopic transport approaches widely tested in
heavy ion collisions 
\cite{chomazPR,baranPR,guarneraPLB373,colonnaNPA642,BaranNPA703}.
The correct treatment of fluctuations is essential to reproduce the 
dynamics of spinodal 
instabilities in the dilute expansion phase. 

We focus on central collisions, $b = 2~fm$, considering symmetric 
reactions
between systems having three different initial asymmetry:
%($\beta = N/Z$):
$^{112}Sn + ^{112}Sn,^{124}Sn + ^{124}Sn,
^{132}Sn + ^{132}Sn,$ with $(N/Z)_{in}$ = 1.24,1.48,1.64, respectively.
The considered
beam energy is 50 MeV/A.
1200 events have been run for each reaction and for each of the two
symmetry energies adopted (asy-soft and asystiff, see before)
\cite{col07}.

In central collisions, after the initial collisional shock, the system
expands and breaks up into many pieces, due to the development of volume
(spinodal) and surface instabilities.  The formation of
a bubble-like configuration is observed, where the initial
fragments are located.

\begin{figure}
%\vspace{0.9cm}
\begin{center}
\includegraphics[width=6.cm]{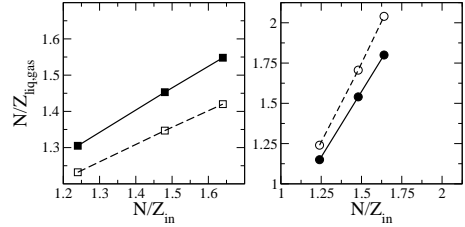}
\caption{The N/Z of the liquid (left) and of the gas (right) phase
is displayed as a function of the system initial N/Z.
Full lines and symbols refer to the asystiff parameterization. Dashed
lines and open symbols are for asysoft.
}
\label{isodist}
\end{center}
\end{figure}

First, let us briefly
recall  some general features concerning the isotopic content of fragments and
emitted nucleons, as obtained with the two considered iso-EOS's.
In the following we will restrict our analysis to fragments with charge
in the range between 3 and 10 (that we call intermediate mass fragments
(IMF's) ).
The average N/Z of emitted nucleons (gas phase) and IMF's
is presented in Fig.\ref{isodist} as a function of the initial $(N/Z)_{in}$
of the three colliding systems.
One observes that, generally,
the gas phase (right panel) is more neutron-rich in the
asysoft case, while IMF's (left panel) are more symmetric. 
%neutron-rich in the asystiff case.
%The difference between the asymmetries of the gas and liquid phases
%%and the one of IMF's
%increases
%with the initial $(N/Z)_{in}$ of the system, and is always larger in the
%asysoft case.
This is due to the larger value of the symmetry energy at low density
for the asysoft parameterization
\cite{BaranNPA703}.
It is interesting to note that, in the asystiff case, due to the
rather low value of the symmetry energy,
%this difference is negative for the proton-rich
%system ($^{112}Sn + ^{112}Sn$).
%In fact, in this case
Coulomb effects dominate and the N/Z
of the gas phase becomes lower than that for IMF's, because protons are
preferentially emitted.
%\section{Correlations}
Now we move to investigate in more detail
correlations between fragment isotopic
content and kinematical properties.
The idea in this investigation
is that fragmentation originates from the break-up of a composite source
that expands with a given velocity field.
Since neutrons and protons experience different forces,
one may expect a different radial flow for the two species.
In this case,  the N/Z composition of the source would not be uniform,
but would depend on the radial distance from the center or mass or,
equivalently, on the local velocity.
This trend should then be reflected in a clear correlation between
isospin content and kinetic energy of the formed $IMF$'s, \cite{col07}.

\vskip -1.0cm
\section{Isospin Transport in Peripheral Collisions at the Fermi Energies}

Now we investigate peripheral collisions between similar systems with 
different 
isospin, specifically collisions of different combinations of $Sn$ isotopes. 
The isospin is used here both as a tracer of the reaction mechanism, as
 well as an 
observable of interest with respect to the iso-EOS. In the prevoius 
sections we have
seen the isospin dynamics in instability regions and fragment formation. 
Here we study 
more directly the isospin transport in binary events.
A neck of density below normal density develops between the two 
heavy residues, the evolution of which is driven by the motion of the 
spectators. 
During this phase isospin is transferred to the neck due to the density 
difference 
between the neck and the residues; this effect is called isospin migration,
 which 
leads to a more neutron-rich neck. In addition in collision systems with
 different asymmetry isospin is transported through neck due to the 
asymmetry gradient
 leading to an equilibration of 
the isospin of the residues, which has been called isospin diffusion. 
Thus in asymmetric
 systems there is  a competion of isospin migration and diffusion.  

The isospin transport can be discussed in a compact way by means of the 
chemical
potentials for protons and neutrons as a function of density $\rho$ and 
isospin 
$I$ \cite{isotr05}. From this the $p/n$ currents can be expressed as
\begin{equation}
{\bf j}_{p/n} = D^{\rho}_{p/n}{\bf \nabla} \rho - D^{I}_{p/n}{\bf \nabla} I
\end{equation}
with $D^{\rho}_{p/n}$ the drift, and
$D^{I}_{p/n}$ the diffusion coefficients for transport, which are given 
explicitely
 in ref. \cite{isotr05}. Of interest here are the differences of currents 
between protons 
and neutrons which have a simple relation to the density dependence of the 
symmetry energy
\begin{eqnarray}
D^{\rho}_{n} - D^{\rho}_{p}  & \propto & 4 I \frac{\partial E_{sym}}
{\partial \rho} \,
 ,  \nonumber\\
D^{I}_{n} - D^{I}_{p} & \propto & 4 \rho E_{sym} \, .
\end{eqnarray}
Thus the isospin transport due to density gradients, i.e. isospin migration, 
depends on the slope of the symmetry energy, or the symmetry pressure, 
while the 
transport due to isospin concentration gradients, i.e. isospin diffusion, 
depends on
 the absolute value of the symmetry energy. In peripheral collisions 
discussed here, 
residues of about normal density are in contact with the neck region of 
density below 
saturation density. In this region of density a stiff iso-EOS has a smaller 
value but 
a larger slope compared to a soft iso-EOS. Correspondingly we expect 
opposite effects of 
these models on the migration and diffusion of isospin.

We will in 
particular discuss the so-called Imbalance Ratio (also called Rami- or 
transport-ratio),
 which is defined as \cite{imbalance}
\begin{equation} 
R^x_{P,T} = \frac{2x^M-(x^H+x^L)}{(x^H-x^L)} .
\label{imb_rat}
\end{equation}
Here, $x$ is an isospin sensitive quantity that is to be investigated with 
respect
 to equilibration, as e.g. the asymmetry $I=(N-Z)/(N+Z)$ which is considered 
here
 primarily, but also other quantities, such as isoscaling coefficients, 
ratios of 
production of light fragments, etc. The indices $H$ and $L$ refer to the 
symmetric 
reaction between the heavy (n-rich) and the light (n-poor) systems, while 
$M$ refers 
to the mixed reaction. $P,T$ denote the rapidity region, in which this 
quantity is 
measured, in particular the projectile and target rapidity regions. 
Clearly, this ratio 
is $\pm1$ in the projectile and target region, respectively, for complete 
transparency, 
and oppositely for complete rebound, while it is zero for complete 
equilibration.

\subsubsection*{Correlation with kinetic energy loss}

The centrality dependence of the Imbalance Ratio, for (Sn,Sn) collisions,
has been investigated in experiments \cite{tsang92} as well as in theory
\cite{isotr05,BALi} with rather promising results for the sensitivity to
the symmetry term stiffness. We propose here a new analysis
 which appears experimentally more selective. 
The interaction time certainly
 influences the amount of  isospin equilibration, see
refs. \cite{isotr05,isotr07}.
On the other hand, longer interaction times should be correlated to
a larger 
dissipation. The dissipation, in turn, has been measured, i.e. in deep 
inelastic 
collisions, by the kinetic energy loss. 

\begin{figure} 
\begin{center}
\includegraphics[width=6.0cm,angle=-90]{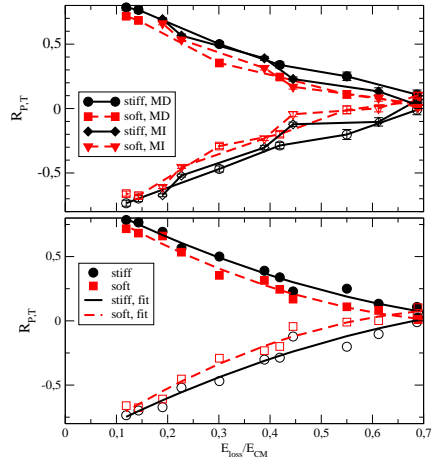}
%\end{picture}
\vspace{0.5cm}
\caption{Imbalance ratio as a function of relative energy loss: (upper panel)
 Separately for stiff (solid) and soft (dashed) iso-EOS, and for MD 
(circles and diamonds) and MI (squares and tringles)interactions, in the 
projectile 
region (full symbols) and the target region (open symbols).
(lower panel) Quadratic fit to all points for the stiff (solid), resp. soft
 (dashed) iso-EOS.
}
\end{center}
\label{imb_eloss}
\end{figure}

In Fig.3 we present the imbalance ratio as a function of 
the
 relative energy loss (H:124Sn, L:112Sn, two beam energies 35 and 50AMeV).
 In the upper panel we separate 
the results for 
Momentum Dependent and Independent (MD vs MI)
interactions and stiff and soft iso-EOS, in each case the results for 
35 and 50 
MeV are collected together and connected by lines. We 
now see that the points for the different incident energies approximately 
fall on one 
line, however with some scatter around it.

The curves for the soft EOS (dashed) 
are generally lower in the projectile region (and oppositely for the target 
region), 
i.e. show more equilibration, that those for the stiff EOS. In order to 
emphasize this 
trend we have, in the lower panel of the figure, collected together the 
values for the 
stiff (circles, solid) and the soft (squares, dashed) iso-EOS, and fitted 
them by a 
quadratic curve. It is seen that this fit gives a good representation of 
the trend of 
the results.

The difference between the curves for the stiff and soft iso-EOS in the 
lower panel
then isolates the effect of the iso-EOS from kinematical effects depending 
on the 
interaction time. It is seen, that there is a systematic effect of the 
symmetry 
energy of the order of about 20 percent, which should be measurable. The 
correlation
 suggested in Fig.3 should represent a general feature of 
isospin 
diffusion, and it would be of great interest to verify it experimentally.

\section{Relativistic Collisions}
 Finally we focus our attention on relativistic heavy ion collisions, that
provide a unique terrestrial opportunity to probe the in-medium nuclear
interaction at high densities. 
An effective Lagrangian approach to the hadron interacting system is
extended to the isospin degree of freedom: within the same frame equilibrium
properties ($EoS$, \cite{qhd}) and transport dynamics 
\cite{KoPRL59,GiessenRPP56} can be consistently derived.
Within a covariant picture of the nuclear mean field, 
 for the description of the symmetry energy at saturation
(a) only the Lorentz vector $\rho$ mesonic field, 
and (b) both, the vector $\rho$ (repulsive) and  scalar 
$\delta$ (attractive) effective 
fields \cite{liu,gait04} can be included. 
In the latter case the competition between scalar and vector fields leads
to a stiffer symmetry term at high density \cite{liu,baranPR}. 
%We present
%here observable effects, in fact enhanced, in the dynamics of heavy ion 
%collisions. 
%Here we focus our attention on collective isospin flows, in 
%particular the elliptic ones, and on the isospin content of particle 
%production,
%in particular kaons. 
%We finally show that in the compression stage of isospin asymmetric collisions
%we can enter a mixed deconfined phase, if the $EoS$ conditions for the
%existence of quark stars are met.
%%%%%%%%%%%%%%%%%%%%%%%%%%%%%%%%%%%%
%\section{Relativistic Transport}
%%%%%%%%%%%%%%%%%%%%%%%%%%%%%%%%%%%%%%%%%%%%%%%%%%%%%%%%%%%%%%%%%%%
%The starting point is
%a simple phenomenological version of the Non-Linear (with respect to the 
%iso-scalar, Lorentz scalar $\sigma$ field) Walecka effective theory 
%which corresponds 
%to the 
%Hartree or Relativistic Mean Field ($RMF$) approximation within the 
%Quantum-Hadro-Dynamics \cite{qhd}. 
%According to this model 
The presence of the hadronic medium leads to effective masses and 
momenta $M^{*}=M+\Sigma_{s}$,   
 $k^{*\mu}=k^{\mu}-\Sigma^{\mu}$, with
$\Sigma_{s},~\Sigma^{\mu}$
 scalar and vector self-energies. 
For asymmetric matter the self-energies are different for protons and 
neutrons, depending on the isovector meson contributions. 
We will call the 
corresponding models as $NL\rho$ and $NL\rho\delta$, respectively, and
just $NL$ the case without isovector interactions. 

For the more general $NL\rho\delta$ case  
the self-energies 
of protons and neutrons read:
%%%%%%%%%%%%%%%
\begin{equation}
\Sigma_{s}(p,n) = - f_{\sigma}\sigma(\rho_{s}) \pm f_{\delta}\rho_{s3}, 
~~~
\Sigma^{\mu}(p,n) = f_{\omega}j^{\mu} \mp f_{\rho}j^{\mu}_{3},~~
(n:~upper~signs),
\label{selfen}
\end{equation}
%%%%%%%%%%%%%%% 
where $\rho_{s}=\rho_{sp}+\rho_{sn},~
j^{\alpha}=j^{\alpha}_{p}+j^{\alpha}_{n},\rho_{s3}=\rho_{sp}-\rho_{sn},
~j^{\alpha}_{3}=j^{\alpha}_{p}-j^{\alpha}_{n}$ are the total and 
isospin scalar 
densities and currents and $f_{\sigma,\omega,\rho,\delta}$  are the coupling 
constants of the various 
mesonic fields. 
$\sigma(\rho_{s})$ is the solution of the non linear 
equation for the $\sigma$ field \cite{liu,baranPR}.
For the description of heavy ion collisions we solve
the covariant transport equation of the Boltzmann type 
 \cite{KoPRL59,GiessenRPP56}  within the 
Relativistic Landau
Vlasov ($RLV$) method, using phase-space Gaussian test particles 
\cite{FuchsNPA589},
and applying
a Monte-Carlo procedure for the hard hadron collisions.
The collision term includes elastic and inelastic processes involving
the production/absorption of the $\Delta(1232 MeV)$ and $N^{*}(1440
MeV)$ resonances as well as their decays into pion channels,
 \cite{FeriniNPA762,Fer06}.
A larger repulsive vector contribution to the neutron energies is given by the
$\rho$-coupling. This is rapidly increasing with density when the $\delta$ 
field is included \cite{liu,baranPR}. As a consequence we expect a good 
sensitivity to the covariant structure of the isovector fields in nucleon 
emission and particle production data. Moreover the presence of a 
{\it Lorentz magnetic} term in the relativistic transport equation
\cite{KoPRL59,GiessenRPP56,baranPR} will enhance the dynamical effects
of vector fields \cite{GrecoPLB562}.

\subsubsection*{Isospin Flows}
(n,p) flow differences will be directly affected.
In ref.\cite{GrecoPLB562}
transverse and elliptic flows results are shown
for 
the $^{132}Sn+^{124}Sn$
reaction at $1.5~AGeV$ ($b=6fm$). 
The effect of the different structure of the 
isovector channel is clear. Particularly evident is the splitting in 
the high $p_t$
region of the elliptic flow.
 In the $(\rho+\delta)$ dynamics the high-$p_t$ neutrons show a much larger 
$squeeze-out$.
This is fully consistent with an early emission (more spectator shadowing)
due to the larger $\rho$-field in the compression stage.

\vskip -1.0cm
\subsubsection*{Isospin effects on sub-threshold kaon 
production at intermediate energies} 
Kaon production has been proven to be a reliable observable for the
high density $EoS$ in the isoscalar sector 
\cite{FuchsPPNP56,HartPRL96}
Here we show that the $K^{0,+}$
production (in particular the $K^0/K^+$ yield ratio) can be also used to 
probe the isovector part of the $EoS$,
\cite{Fer06,Pra07}.

Using our $RMF$ transport approach  we have analyzed 
pion and kaon production in central $^{197}Au+^{197}Au$ collisions in 
the $0.8-1.8~AGeV$
 beam 
energy range, comparing models giving the same ``soft'' $EoS$ for symmetric 
matter and with different effective field choices for 
$E_{sym}$. We will use three Lagrangians with constant 
nucleon-meson 
couplings ($NL...$ type, see before) and one with density
dependent couplings ($DDF$, see \cite{gait04}), recently suggested 
for better nucleonic properties of neutron stars \cite{Klahn06,Liubo07}.
%In the $DDF$ model
%the $f_{\rho}$ is exponentially decreasing with density, resulting in a 
%rather "soft" 
%symmetry term at high density. 
%The hadron mean field propagation, which goes beyond the 
%``collision cascade'' picture, is
%essential for particle production yields: in particular the
%isospin dependence of the self-energies directly affects the
%energy balance of the inelastic channels.

\begin{figure}[t] 
\begin{center}
\includegraphics[scale=0.27]{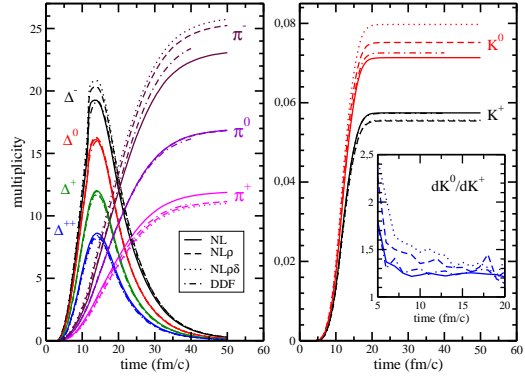} 
%\vskip -1.0cm
\caption{\small{Time evolution of the $\Delta^{\pm,0,++}$ resonances
and pions $\pi^{\pm,0}$ 
(left),  and  kaons ($K^{+,0}$
 (right) for a central ($b=0$ fm impact parameter)  
Au+Au collision at 1 AGeV incident energy. Transport calculation using the  
$NL, NL\rho, NL\rho\delta$ and $DDF$ models for the iso-vector part of the  
nuclear $EoS$ are shown. The inset contains the differential $K^0/K^+$  ratio
as a function of the kaon emission time.  
}}
\vskip -0.5cm
\label{kaon1} 
\end{center}
\end{figure}

Fig. \ref{kaon1} reports  the temporal evolution of $\Delta^{\pm,0,++}$  
resonances, pions ($\pi^{\pm,0}$) and kaons ($K^{+,0}$)  
for central Au+Au collisions at $1AGeV$.
%%%%%%%%%%%%%%%%%%%%%%%%%%%%%%%%%%%%%%%%%%%%%%%%%%%%%%%%%%%%%%%%%%% 
%%%%%%%%%%%%%%%%%%%%%%%%%%%%%%%%%%%%%%%%%%%%%%%%%%%%%%%%%%%%%%%%%%%%%%%%%%%%% 
It is clear that, while the pion yield freezes out at times of the order of 
$50 fm/c$, i.e. at the final stage of the reaction (and at low densities),
kaon production occur within the very early (compression) stage,
 and the yield saturates at around $20 fm/c$. 
 Consistently, as shown in the
insert, larger isospin effects are expected for emitted kaons.

When isovector fields are included the symmetry potential energy in 
neutron-rich matter is repulsive for neutrons and attractive for protons.
In a $HIC$ this leads to a fast, pre-equilibrium, emission of neutrons.
 Such a $mean~field$ mechanism, often referred to as isospin fractionation
\cite{bao,baranPR}, is responsible for a reduction of the neutron
to proton ratio during the high density phase, with direct consequences
on particle production in inelastic $NN$ collisions.

$Threshold$ effects represent a more subtle point. The energy 
conservation in
a hadron collision in general has to be formulated in terms of the canonical
momenta, i.e. for a reaction $1+2 \rightarrow 3+4$ as
$
s_{in} = (k_1^\mu + k_2^\mu)^2 = (k_3^\mu + k_4^\mu)^2 = s_{out}.
$
Since hadrons are propagating with effective (kinetic) momenta and masses,
 an equivalent relation should be formulated starting from the effective
in-medium quantities $k^{*\mu}=k^\mu-\Sigma^\mu$ and $m^*=m+\Sigma_s$, where
$\Sigma_s$ and $\Sigma^\mu$ are the scalar and vector self-energies,
Eqs.(\ref{selfen}).
The self-energy contributions will influence the particle production at the
level of thresholds as well as of the phase space available in the final 
channel.

In the few AGeV region
the effect is larger for the $K^{0}/K^{+}$ compared to the $\pi^-/\pi^+$
ratio. This is due to the subthreshold production and to the fact that
the isospin effect enters twice in the two-step production of kaons, see
\cite{Fer06}. 
Interestingly the Iso-$EoS$ effect for pions is increasing at lower energies,
when approaching the production threshold.

%In neutron-rich colliding systems {\it Mean field} 
%and {\it threshold} effects
%are acting in opposite directions on particle production  and might 
%compensate each other.
% As an example, $nn$
%collisions excite $\Delta^{-,0}$ resonances which decay mainly to $\pi^-$.
% In a
%neutron-rich matter the mean field effect pushes out neutrons making the 
%matter more symmetric and thus decreasing the $\pi^-$ yield. The threshold 
%effect on the other hand is increasing the rate of $\pi^-$'s due to the
%enhanced production of the $\Delta^-$ resonances: 
%now the $nn \rightarrow p\Delta^-$ process is favored
%(with respect to $pp \rightarrow n\Delta^{++}$) 
% since more effectively a neutron is converted into a proton.
%Such interplay between the two mechanisms cannot be 
%fully included in a non-relativistic dynamics,
%in particular in calculations where the baryon symmetry potential is
%treated classically \cite{BaoPRC71,QLiPRC72}.

We have to note that in a study of kaon production in excited nuclear
matter the dependence of the $K^{0}/K^{+}$ yield ratio on the effective
isovector interaction appears much larger (see Fig.8 of 
ref.\cite{FeriniNPA762}).
The point is that in the non-equilibrium case of a heavy ion collision
the asymmetry of the source where kaons are produced is in fact reduced
by the $n \rightarrow p$ ``transformation'', due to the favored 
$nn \rightarrow p\Delta^-$ processes. This effect is almost absent at 
equilibrium due to the inverse transitions, see Fig.3 of 
ref.\cite{FeriniNPA762}. Moreover in infinite nuclear matter even the fast
neutron emission is not present. 
This result clearly shows that chemical equilibrium models can lead to
uncorrect results when used for transient states of an $open$ system.
\vskip -1.0cm
\section{Testing Deconfinement at High Isospin Density}
The hadronic matter is expected to undergo a phase transition 
into a deconfined phase of quarks and gluons at large densities 
and/or high temperatures. On very general grounds,
the transition's critical densities are expected to depend
on the isospin of the system, but no experimental tests of this 
dependence have been performed so far.

Here we suggest the possibility of an $earlier$ transition density
to a mixed phase in isospin asymmetric systems \cite{deconf06}. 
% \cite{MuellerNPA618} in a systematic way,
% exploring also
%the possibility of forming a mixed-phase of quarks and hadrons in
%experiments at energies of the order of a few $GeV$ per nucleon.
Concerning the hadronic phase, we use the relativistic
non-linear model of Glendenning-Moszkowski (in particular the ``soft''
$GM3$ choice) 
\cite{GlendenningPRL18}, where the isovector part is treated 
just with $\rho$ meson coupling, and
the iso-stiffer $NL\rho\delta$ interaction. 
For the quark phase we consider the $MIT$ bag model 
%\cite{MitbagPRD9}
with various bag pressure constants.  In particular 
we are interested in those parameter sets
which would allow the existence of quark stars
\cite{DragoPLB511}, i.e. parameters sets for
which the so-called Witten-Bodmer hypothesis is satisfied
\cite{WittenPRD30,BodmerPRD4}. 
One of the
aim of our work it to show that if quark stars are indeed possible,
it is then very likely to find signals of the formation of a mixed
quark-hadron phase in intermediate-energy heavy-ion experiments
\cite{deconf06}.

The structure of the mixed phase is obtained by
imposing the Gibbs conditions \cite{Landaustat,GlendenningPRD46} for
chemical potentials and pressure and by requiring
the conservation of the total baryon and isospin densities
\begin{eqnarray}\label{gibbs}
&&\mu_B^{(H)} = \mu_B^{(Q)}\, ,~~  
\mu_3^{(H)} = \mu_3^{(Q)} \, ,  \nonumber \\
&&P^{(H)}(T,\mu_{B,3}^{(H)}) = P^{(Q)} (T,\mu_{B,3}^{(Q)})\, ,\nonumber \\
&&\rho_B=(1-\chi)\rho_B^H+\chi\rho_B^Q \, , \nonumber \\
&&\rho_3=(1-\chi)\rho_3^H+\chi\rho_3^Q\, , 
\end{eqnarray}
where $\chi$ is the fraction of quark matter in the mixed phase.
\begin{figure}
\begin{center}
\includegraphics[angle=+90,scale=0.28]{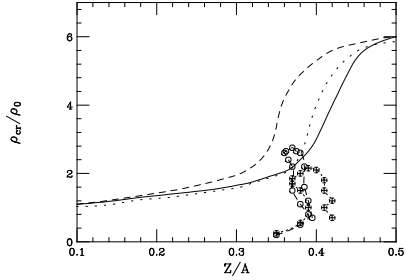}
\vskip -0.3cm
\caption{
\small{Variation of the transition density with proton fraction for various
hadronic $EoS$ parameterizations. Dotted line: $GM3$ parametrization;
 dashed line: $NL\rho$ parametrization; solid line: $NL\rho\delta$ 
parametrization. For the quark $EoS$, the $MIT$ bag model with
$B^{1/4}$=150 $MeV$.
The points represent the path followed
in the interaction zone during a semi-central $^{132}$Sn+$^{132}$Sn
collision at $1~AGeV$ (circles) and at $300~AMeV$ (crosses). 
}}
\vskip -0.5cm
\label{rhodelta}
\end{center}
\end{figure}
In this way we get the $binodal$ surface which gives the phase coexistence 
region
in the $(T,\rho_B,\rho_3)$ space.
%\cite{GlendenningPRD46,MuellerNPA618}. 
For a fixed value of the
conserved charge $\rho_3$ 
 we will study the boundaries of the mixed phase
region in the $(T,\rho_B)$ plane. 
In the hadronic phase the charge chemical potential is given by
$
\mu_3 = 2 E_{sym}(\rho_B) \frac{\rho_3}{\rho_B}\, .
$ 
Thus, we expect critical densities
rather sensitive to the isovector channel in the hadronic $EoS$.

In Fig.~\ref{rhodelta}  we show the crossing
density $\rho_{cr}$ separating nuclear matter from the quark-nucleon
mixed phase, as a function of the proton fraction $Z/A$.  
We can see the effect of the
$\delta$-coupling towards an $earlier$ crossing due to the larger
symmetry repulsion at high baryon densities.
In the same figure we report the paths in the $(\rho,Z/A)$
plane followed in the c.m. region during the collision of the n-rich
 $^{132}$Sn+$^{132}$Sn system, at different energies. At
$300~AMeV$ we are just reaching the border of the mixed phase, and we are
well inside it at $1~AGeV$. 
%fluctuations could help in  reducing the density at which drops of quark
%matter form. The reason is that a small bubble can
%be energetically favored if it contains quarks whose Z/A ratio is
%{\it smaller} than the average value of the surrounding region. This
%is again due to the strong Z/A dependence of the free energy, which
%favors clusters having a small electric charge. 
%Moreover, since 
%fluctuations favor the formation of bubbles having a smaller Z/A,
%neutron emission from the central collision area should be suppressed,
%which could give origin to specific signatures of the mechanism
%described in this paper. 
We expect a {\it neutron trapping}
effect, supported by statistical fluctuations as well as by a 
symmetry energy difference in the
two phases.
In fact while in the hadron phase we have a large neutron
potential repulsion (in particular in the $NL\rho\delta$ case), in the
quark phase we only have the much smaller kinetic contribution.
% If in a
%pure hadronic phase neutrons are quickly emitted or ``transformed'' in
%protons by inelastic collisions, when the mixed phase
%starts forming, neutrons are kept in the interacting system up to the
%subsequent hadronization in the expansion stage \cite{deconf06}.
Observables related to such neutron ``trapping'' could be an
inversion in the trend of the formation of neutron rich fragments
and/or of the $\pi^-/\pi^+$, $K^0/K^+$ yield ratios for reaction
products coming from high density regions, i.e. with large transverse
momenta.  In general we would expect a modification of the rapidity
distribution of the emitted ``isospin'', with an enhancement at
mid-rapidity joint to large event by event fluctuations..

\vskip -1.0cm
\section{Perspectives}
We have shown that {\it violent} collisions of n-rich heavy ions 
from low to relativistic energies
can bring new information on the isovector part of the in-medium interaction, 
qualitatively different from equilibrium
$EoS$ properties. We have presented quantitative results 
in a wide range of beam energies.
At low energies we see isospin effects on the dissipation in deep 
inelastic collisions, at Fermi energies the Iso-EoS sensitivity of the isospin 
transport in fragment reactions and finally at intermediate the dependence of
differential flows on the $Iso-MD$ and effective mass splitting. 
In relativistic collisions we have shown the possibility of a direct 
$measure$ of the Lorentz structure of the isovector fields at high baryon 
density. We have presented quantitative results for differential
collective flows and yields of charged pion and kaon ratios.
Important non-equilibrium effects for particle production are stressed.
Finally our study supports the possibility of observing
precursor signals of the phase transition to a mixed hadron-quark matter
at high baryon density in the collision, central or semi-central, of
neutron-rich heavy ions in the energy range of a few $GeV$ per
nucleon.  
In conclusion the results presented in this "Isospin Journey" appear 
very promising for 
the possibility of exciting new results from dissipative collisions
with radioactive beams.

 \vskip 0.2cm
\noindent
{\it Acknowledgements}.
We warmly thanks A.Drago and A.Lavagno for the intense 
collaboration on the
mixed hadron-quark phase transition at high baryon and isospin density.
                      
                                    %
%%%%%%%%%%%%%%%%%%%%%%%%%%%%%%%%%%%%%%%%%%%%%%%%%%%%%%%%%%%%%%%%%%%%%%%%%

%\include{nn06_ditorobib}

%%%%%%%%%%%%%%%%%%%%%%%%%%%%%%%%%%%%%%%%%%%%%%%%%%%%%%%%%%%%%%

%%%%%%%%%%%%%%%%%%%%%%%%%%%%%%%%%%%%%%%%%%%%%%%%%%%%%%%%%%%%%%
\end{document}